\documentclass{article}

\usepackage{complexity}
\usepackage{amsthm}
\usepackage{thmtools}
\usepackage[
  margin=1.5cm,
  includefoot,
  footskip=30pt,
]{geometry}

\usepackage{algorithm,algorithmicx}

\usepackage{dirtytalk}
\usepackage{bbold}

\usepackage{enumerate} 

\usepackage{amsmath} 
\usepackage{amssymb}

\usepackage[mathletters]{ucs}

\usepackage{amsmath,amstext,amssymb,amsfonts}
\usepackage{graphicx}
\usepackage{xcolor}
\usepackage{tabularx}
\usepackage{verbatim}
\usepackage{bbm}

\usepackage{microtype}

\newtheorem{theorem}{Theorem}
\newtheorem*{theorem*}{Theorem}

\newtheorem*{claim*}{Claim}

\newtheorem{proposition}[theorem]{Proposition}
\newtheorem*{proposition*}{Proposition}
\newtheorem{lemma}[theorem]{Lemma}
\newtheorem*{lemma*}{Lemma}

\newtheorem*{conjecture*}{Conjecture}

\newtheorem*{fact*}{Fact}

\newtheorem*{hypothesis*}{Hypothesis}

\theoremstyle{definition}
\newtheorem{definition}[theorem]{Definition}

\usepackage{prettyref}
\newcommand{\savehyperref}[2]{\texorpdfstring{\hyperref[#1]{#2}}{#2}}

\newrefformat{parta}{\savehyperref{#1}{a}}
\newrefformat{partb}{\savehyperref{#1}{b}}
\newrefformat{eq}{\savehyperref{#1}{\textup{(\ref*{#1})}}}
\newrefformat{lem}{\savehyperref{#1}{Lemma~\ref*{#1}}}
\newrefformat{def}{\savehyperref{#1}{Definition~\ref*{#1}}}
\newrefformat{thm}{\savehyperref{#1}{Theorem~\ref*{#1}}}
\newrefformat{cor}{\savehyperref{#1}{Corollary~\ref*{#1}}}
\newrefformat{cha}{\savehyperref{#1}{Chapter~\ref*{#1}}}
\newrefformat{sec}{\savehyperref{#1}{Section~\ref*{#1}}}
\newrefformat{app}{\savehyperref{#1}{Appendix~\ref*{#1}}}
\newrefformat{tab}{\savehyperref{#1}{Table~\ref*{#1}}}
\newrefformat{fig}{\savehyperref{#1}{Figure~\ref*{#1}}}
\newrefformat{hyp}{\savehyperref{#1}{Hypothesis~\ref*{#1}}}
\newrefformat{alg}{\savehyperref{#1}{Algorithm~\ref*{#1}}}
\newrefformat{sdp}{\savehyperref{#1}{SDP~\ref*{#1}}}
\newrefformat{qp}{\savehyperref{#1}{QP~\ref*{#1}}}
\newrefformat{vp}{\savehyperref{#1}{VP~\ref*{#1}}}
\newrefformat{lp}{\savehyperref{#1}{LP~\ref*{#1}}}
\newrefformat{rem}{\savehyperref{#1}{Remark~\ref*{#1}}}
\newrefformat{item}{\savehyperref{#1}{Item~\ref*{#1}}}
\newrefformat{step}{\savehyperref{#1}{step~\ref*{#1}}}
\newrefformat{conj}{\savehyperref{#1}{Conjecture~\ref*{#1}}}
\newrefformat{fact}{\savehyperref{#1}{Fact~\ref*{#1}}}
\newrefformat{prop}{\savehyperref{#1}{Proposition~\ref*{#1}}}
\newrefformat{claim}{\savehyperref{#1}{Claim~\ref*{#1}}}
\newrefformat{relax}{\savehyperref{#1}{Relaxation~\ref*{#1}}}
\newrefformat{red}{\savehyperref{#1}{Reduction~\ref*{#1}}}
\newrefformat{part}{\savehyperref{#1}{Part~\ref*{#1}}}
\newrefformat{prob}{\savehyperref{#1}{Problem~\ref*{#1}}}
\newrefformat{ass}{\savehyperref{#1}{Assumption~\ref*{#1}}}
\newrefformat{cons}{\savehyperref{#1}{Construction~\ref*{#1}}}
\newrefformat{obs}{\savehyperref{#1}{Observation~\ref*{#1}}}

\newcommand{\Sref}[1]{\hyperref[#1]{\S\ref*{#1}}}

\renewcommand{\leq}{\leqslant}

\renewcommand{\geq}{\geqslant}

\usepackage{bm}

\newcommand{\set}[1]{\left\{#1\right\}}

\newcommand{\opt}{{\sf OPT}}

\definecolor{DSgray}{cmyk}{0,0,0,0.7}

\usepackage[pdftex,pagebackref,colorlinks,linkcolor=cyan, filecolor = blue, citecolor = cyan, urlcolor  = teal]{hyperref}

    \definecolor{urlcolor}{rgb}{0,.145,.698}
    \definecolor{linkcolor}{rgb}{.71,0.21,0.01}
    \definecolor{citecolor}{rgb}{.12,.54,.11}

    \definecolor{ansi-black}{HTML}{3E424D}
    \definecolor{ansi-black-intense}{HTML}{282C36}
    \definecolor{ansi-red}{HTML}{E75C58}
    \definecolor{ansi-red-intense}{HTML}{B22B31}
    \definecolor{ansi-green}{HTML}{00A250}
    \definecolor{ansi-green-intense}{HTML}{007427}
    \definecolor{ansi-yellow}{HTML}{DDB62B}
    \definecolor{ansi-yellow-intense}{HTML}{B27D12}
    \definecolor{ansi-blue}{HTML}{208FFB}
    \definecolor{ansi-blue-intense}{HTML}{0065CA}
    \definecolor{ansi-magenta}{HTML}{D160C4}
    \definecolor{ansi-magenta-intense}{HTML}{A03196}
    \definecolor{ansi-cyan}{HTML}{60C6C8}
    \definecolor{ansi-cyan-intense}{HTML}{258F8F}
    \definecolor{ansi-white}{HTML}{C5C1B4}
    \definecolor{ansi-white-intense}{HTML}{A1A6B2}

\usepackage{fourier}

\usepackage{tikz}
\usepackage{caption}
\usepackage{subcaption}

\usepackage{hyperref}
\usepackage{amsthm}
\usepackage{amsmath}
\usepackage{amsfonts}
\usepackage{cleveref}
\usepackage{thmtools, thm-restate}

\title{Improved Hardness of Approximation for Geometric Bin Packing} %TODO Please add

\author{
    Arka Ray\footnote{This research was supported in part by Anand Louis's Pratiksha Trust Young Investigator Award and in part by Rahul Saladi's Pratiksha Trust Young Investigator Award. Arka Ray is currently supported by the Walmart Center for Tech Excellence at IISc (CSR Grant WMGT-23-0001).}\\
    Indian Institute of Science, Bengaluru\\
    \href{mailto:arkaray@iisc.ac.in}{arkaray@iisc.ac.in}
    \and
    Sai Sandeep\footnote{Research supported in part by NSF grant CCF-2228287.}\\
    UC Berkeley, CA, USA\\
    \href{mailto:saisandeep@berkeley.edu}{saisandeep@berkeley.edu}
}
    
\date{}

\newcommand{\gpd}{\textsf{gpd}}
\newcommand{\pdim}{\textsf{pdim}}
\newcommand{\F}{\mathcal F}

\renewcommand{\epsilon}{\varepsilon}

\begin{document}

\maketitle

\begin{abstract}
The Geometric Bin Packing (GBP) problem is a generalization of Bin Packing where the input is a set of $d$-dimensional rectangles, and the goal is to pack them into $d$-dimensional unit cubes efficiently. It is NP-hard to obtain a PTAS for the problem, even when $d=2$. For general $d$, the best-known approximation algorithm has an approximation guarantee that is exponential in $d$.
In contrast, the best hardness of approximation is still a small constant inapproximability from the case when $d=2$. In this paper, we show that the problem cannot be approximated within a $d^{1-\epsilon}$ factor unless $\NP=\P$. 

%Our reduction outputs instances where the dimension $d$ grows with the number of rectangles.
Recently, $d$-dimensional Vector Bin Packing, a problem closely related to the GBP, was shown to be hard to approximate within a $\Omega(\log d)$ factor when $d$ is a fixed constant, using a notion of Packing Dimension of set families.
In this paper, we introduce a geometric analog of it, the Geometric Packing Dimension of set families.  
While we fall short of obtaining similar inapproximability results for the Geometric Bin Packing problem when $d$ is fixed, we prove a couple of key properties of the Geometric Packing Dimension which highlight fundamental differences between Geometric Bin Packing and Vector Bin Packing. 
%, showing that it is significantly richer than the Packing Dimension of set families. 
\end{abstract}

%TODO mandatory: add short abstract of the document

%\newpage
\section{Introduction}
\label{sec:intro}

In the Geometric Bin Packing (GBP) problem, the input is a set of $d$-dimensional rectangles, and the objective is to pack them %\footnote{In this work, we study the case where rotations are not allowed and GBP will refer to this case unless stated otherwise.}
into a minimum number of $d$-dimensional unit cubes.
In this work, we study the variant where rotations are not allowed, and GBP will refer to this case unless stated otherwise.
The problem is widely applicable in practice and has received a lot of attention in the approximation algorithms community (see~\cite{ChristensenKPT17} for detailed discussion).
It is one of the two most extensively studied generalizations of Bin Packing (the other being Vector Bin Packing), which corresponds to the case when $d=1$.
Bin Packing is a classical NP-hard problem and has an asymptotic\footnote{The asymptotic approximation ratio of an algorithm corresponds to the case when the optimal value is large enough. We shall simply say approximation ratio to refer to asymptotic approximation ratio hereon and specify explicitly when absolute approximation ratio is used.} PTAS~\cite{VegaL81}. 
Already for $d=2$, i.e., the problem of packing (2-dimensional) rectangles in square boxes, a PTAS cannot be obtained unless $\P=\NP$.
 The best polynomial-time algorithm known for this 2-dimensional setting is by Bansal and Khan~\cite{BansalK14} with an approximation ratio of 1.406, and the best hardness result is $1+\frac{1}{2196}$ due to Chlebik and Chleb{\'i}kov{\'a}~\cite{CC09}. 

Typically, multidimensional packing problems have two variants: when the dimension $d$ is part of the input size and when $d$ is a fixed constant independent of the input.
The key difference is that when $d$ is fixed, the algorithms are allowed to run in time $n^{f(d)}$ for an arbitrary function $f$.
On the algorithmic side, even with fixed $d$, the best-known algorithm has an approximation ratio
$T_{\infty}^{d-1}$~\cite{Caprara02} where $T_\infty$ denotes the Harmonic constant \footnote{$T_\infty$ is defined by
$T_\infty = \sum_{i=1}^\infty \frac{1}{t_i-1}$ with $t_1=2$ and $t_{i+1}=t_i(t_i-1)+1$.}, whereas the hardness is still the non-existence of PTAS from the $2$-dimensional case.
For the GBP, for both cases, previously, no hardness of approximation growing with $d$ was known, and this has been one of the ten open problems in a recent survey on multidimensional packing problems~\cite{ChristensenKPT17}.
Note that obtaining hardness for the case when $d$ is part of the input is easier than showing hardness for the fixed $d$ case.

In this work, we obtain a strong inapproximability result for GBP when $d$ is part of the input. 
\begin{restatable}{theorem}{main}
\label{thm:main_informal}
Geometric Bin Packing is hard to approximate within a $n^{1-\epsilon}$ factor for instances with $n$ items in $n$ dimensions, for every $\epsilon>0$, unless $\NP=\P$. 
\end{restatable}

Our proof is a direct reduction from graph coloring, similar to the proof of hardness of approximation for $d$-dimensional Vector Bin Packing due to Chekuri and Khanna~\cite{CK04}. 

Vector Bin Packing is another generalization of Bin Packing where we are given a set of $d$-dimensional vectors from $[0,1]^d$, and the aim is to partition the set into the minimum number of bins such that for each bin $B$, each component of the sum of the vectors in $B$ is at most 1.
When $d$ is a fixed constant, in a recent work~\cite{Sandeep21}, $\Omega(\log d)$ hardness of approximation for Vector Bin Packing has been obtained.
This result is obtained by a reduction from the Set Cover problem via the \textit{packing dimension} of set systems, which is the minimum dimension in which the set cover of a set system can be embedded as a Vector Bin Packing problem instance. 
In the Set Cover problem, the input is a set family $\mathcal{F}$ on a universe $U$, and the objective is to find the minimum number of sets from $\mathcal{F}$ whose union is $U$.
An instance of the Vector Bin Packing problem can be formulated as an instance of the Set Cover problem.
First, we take a set of items as our universe.
Now, the collection of sets of vectors that fit in a unit cube (each such set is called a ``configuration'' in the bin packing literature) constitutes our instance, and the objective is to find the minimum number of sets~(configurations) whose union covers all the vectors. The inapproximability result for the problem is obtained by reversing this reduction, i.e., by embedding the items in a Set Cover instance as a vector in another Vector Bin Packing instance such that the configurations in the Vector Bin Packing instance are exactly the sets in the given Set Cover instance.
The minimum dimension of the Vector Bin Packing instance that we can output in this way, starting with a Set Cover instance of a set family $\F$, is precisely the packing dimension of the set family $\F$.
The packing dimension of $\F$ is denoted by $\pdim(\F)$.

In this work, we study the analogous notion for the Geometric Bin Packing problem. In particular, we define the \emph{geometric packing dimension} $\gpd(\F)$ of a set system $\F \subseteq 2^U$ on a universe $U$ to be the smallest integer $d$ such that there is an embedding of the elements of $\F$ to $d$-dimensional rectangles such that a set $S $ of elements is in $\F$ if and only if the corresponding rectangles fit in a $d$-dimensional unit cube. If no such an embedding exists, then we say that $\F$ has no finite $\gpd$. 
By obtaining such an embedding in polynomial time, we can get a direct reduction from the Set Cover problem on $\F$ to a GBP instance with dimension $\gpd(\F)$. 
The goal is to find set families that have small packing dimension where Set Cover is hard, thus implying a hardness of approximation for the GBP problem when $d$ is constant.
In this work, while we fall short of this objective, we formally introduce and study $\gpd$ of set families, and prove a couple of properties of it.

For a set family $\F$ on a universe $U$, $\gpd(\F)$ being finite is equivalent to the fact that there exists a Geometric Bin Packing instance where the rectangles correspond to the elements of $U$, and the configurations are exactly the sets in $\F$. It is an interesting question, then, to characterize which set families $\F$ have finite $\gpd(\F)$. 
Similar to the packing dimension~\cite{Sandeep21}, two conditions are necessary for a set system $\F$ to have a finite $\gpd$: first, the set system should be downward closed, i.e., for every $S \in \F$ and $T \subseteq S$, we have $T \in \F$ as well. Second, the set system should not have any isolated elements, i.e., for every $i \in U$, there is $S \in \F$ such that $i \in S$.
In~\cite{Sandeep21}, the author proves that these two conditions are sufficient for a set system to have a finite packing dimension using a very simple embedding.
In particular, they show \autoref{prop:vdim-finite}.

\begin{proposition}[\cite{Sandeep21}]
\label{prop:vdim-finite}
For every set system $\F\subseteq 2^{V}$ that is downward closed, we have $\pdim(\F)\leq 2^{|V|}$.
\end{proposition}

In stark contrast, we show that this does not hold for the geometric packing dimension. 

\begin{restatable}{theorem}{gpdinfinite}
\label{thm:gpd-infinite}
	There is a downward closed set system $\F$ such that $\gpd(\F)$ is not finite. 
\end{restatable}

Our construction is obtained using a set system defined via lines in $\mathbb{F}_3^n$ ($\mathbb F_3$ denotes the field with characteristic 3). The key property that we use is that while the set system is dense enough, every pair of elements appears in exactly one set. 
%For the Geometric Bin Packing problem,~\Cref{thm:gpd-infinite} shows that the configurations of the Geometric Bin Packing cannot be arbitrary set families that are downward closed and have no isolated elements. 

Although \autoref{thm:gpd-infinite} shows that not every downward closed Set Cover instance can be succinctly represented using Geometric Bin Packing instance, one may still hope to reduce some structured set families on which Set Cover is hard to Geometric Bin Packing.
%As mentioned earlier, the main motivation behind defining the Geometric Packing Dimension is to find set systems $\F$ with small $\gpd(\F)$, yet set cover is hard on them. 
%The classical hardness result for the set cover problem~\cite{Feige98} shows that the set cover problem on $n$ elements is hard to approximate within $\Omega(\ln n)$ factor. However, since we are operating in the fixed $d$ setting, and there is already $O(1)$ factor approximation algorithm for GBP in this setting, the set families of~\cite{Feige98} necessarily have $\gpd$ growing with the input size.
In fact, we not only need the $\gpd$ to be finite but also % to let $\gpd$
be a fixed constant independent of the input size for a hardness of approximation result when $d$ is fixed. %, yet Set Cover is hard to approximate on them.
A natural candidate for this are the $(k,B)$-bounded set systems, where each set has size at most $k$, and each element appears in at most $B$ sets, with $k$ and $B$ being fixed constants. Indeed, for the $d$-dimensional Vector Bin Packing, these set systems were used to show the inapproximability result in~\cite{Sandeep21}. 
In particular, they show \autoref{thm:bound-pd}.

\begin{restatable}[\cite{Sandeep21}]{theorem}{bound-pd}
\label{thm:bound-pd}
Let $\F\subseteq 2^{V}$ be a set system on a universe $V$ where each set has cardinality at most $k \geq 2$ and each element appears in at most $B$ sets.
Suppose that $\F$ is simple, i.e., $|S_1\cap S_2|\leq 1$ for any two distinct $S_1,S_2\in \F$.
Also, suppose $\F$ covers $V$, i.e., $\bigcup_{S\in \F}S = V$.
Then,
\begin{align*}
\pdim(\F^{\downarrow})\leq (kB)^{O(1)}
\end{align*}
where $\F^{\downarrow}$ is the downward closure of $\F$, i.e., $\F^{\downarrow}=\set{S\subseteq V| S\subseteq T \text{ for some }T\in \F}$.
\end{restatable}

%\begin{remark*}
%Note that if $\F$ has cardinality at most $k$, each element appears in at most $B$ sets, and $\F$ is simple, then the same is true for $\F^\downarrow$.
%Moreover, if $\F$ is downward closed then $\F^\downarrow = \F$.
%\end{remark*}

However, we show that any bounded set system has a large $\gpd$.

\begin{restatable}{theorem}{bounded}
\label{thm:gpd-simple}
	Let $\F \subseteq 2^U$ be a set family that is $(k,B)$-bounded with $k,B$ constants and has no isolated elements. Then, either $\gpd(\F)$ is not finite, or it is at least $\Omega(|U|)$. 
\end{restatable}

This rules out any direct reduction from Set Cover instances that are bounded to the GBP problem.

Our proof of \autoref{thm:gpd-simple} is obtained by studying the induced matching in such set families. The bounded set systems have a large induced matching, which implies that $\gpd$ has to be large as well. 

%This is a significant bottleneck, since most set cover hardness of approximation results, starting with 

\subsection{Related Work}
Bin Packing is a classical NP-complete problem.
%A simple reduction from Partition Problem shows that an absolute approximation ratio better than $3/2$ is not possible (folklore).
%Hence, most results concerning Bin Packing are stated in terms of asymptotic approximation ratio.
Fernandez de la Vega and Lueker~\cite{VegaL81} used the linear grouping technique to obtain a PTAS.
The best algorithm known has an additive error of $O(\log \opt)$ due to Hoberg and Rothvo{\ss}~\cite{HR17}.
It is still an open problem to determine whether an additive error of 1 is possible or not.

For Geometric Bin Packing Problem when $d=2$, some of the recent work include the $T_\infty+\epsilon$ approximation~\cite{Caprara02}.
Bansal, Caprara, and Sviridenko~\cite{BCS09} improved it further using their Round and Approx framework to obtain a $1+\ln(T_\infty)\approx 1.52$ approximation.
Finally, Bansal and Khan~\cite{BansalK14} gave the $1+\ln(1.5)\approx 1.406$ approximation by showing the Round and Approx framework applies to the $1.5+\epsilon$ approximation due to Jansen and Pr{\"a}del~\cite{JP16}.
When $d>2$, the $T_\infty^{d-1}$-approximation by Caprara~\cite{Caprara02} stands as the current best.
On the hardness side, Bansal et~al.~\cite{BCKS06} showed that there is no PTAS even for $d=2$, unless $\P=\NP$.
This was later improved by Chlebik and Chleb{\'i}kov{\'a}~\cite{CC09} to $1+\frac{1}{2196}$ by modifying the construction in \cite{BCKS06}. 
As stated earlier, the $1+\frac{1}{2196}$ bound is the best hardness result known, even for higher dimensions.

Vector Bin Packing is another well-studied generalization of the Bin Packing Problem.
When $d$ is part of the input, there is a $(d+\epsilon)$-approximation due to Fernandez de la Vega and Luekar~\cite{VegaL81}.
On the hardness side, a simple modification to the reduction by Chekuri and Khanna~\cite{CK04} due to Jan Vondr{\'a}k (see footnote 2 in \cite{BEK06})  gives a $d^{1-\epsilon}$ hardness.
When $d$ is not part of the input the barrier of $d$ was broken by Chekuri and Khanna~\cite{CK04} by giving $\ln d + 2 + \gamma$ appoximation where $\gamma$ denotes the Euler-Mascheroni constant\footnote{$\gamma$ is defined by $\gamma=\lim_{n\to \infty} \sum_{k=1}^{n}\frac{1}{k}-\ln n$.}.
This was improved to $1 + \ln d$, and then to $1 + \ln d - \chi(d)$ by Bansal, Caprara and Sviridenko~\cite{BCS09} and Kulik, Mnich, and Shachnai~\cite{KMS22}\footnote{Kulik, Mnich, and Shachnai~\cite{KMS22} also show a flaw in the analysis of $(\ln(d+1) + 0.807 + \epsilon$)-approximation due to Bansal, Eli\'a\v{s}, and Khan~\cite{BEK06}.}, respectively.
Recently, Sandeep~\cite{Sandeep21} improved the lower bound to $\Omega(\ln d)$ from $1+\frac{1}{599}$ due to Ray~\cite{Ray24} and Woeginger~\cite{Woeginger97}.

For a more comprehensive review of the recent works on approximation algorithms for Bin Packing and related problems, we refer the reader to the survey by Christensen et~al.~\cite{ChristensenKPT17}.

\subsection{Preliminaries} 
\paragraph{Notations} We use $[n]$ to denote $\{1,2,\ldots,n\}$. A set family or set system $\mathcal{F} \subseteq 2^U$ is a family of subsets of $U$. We use boldface letters to denote $d$-dimensional vectors or rectangles. For a $d$-dimensional rectangle $\mathbf u$, we use $u_i$ to denote the $i$th coordinate, and for a $d$-dimensional rectangle $\mathbf v_i$, we use $v_{i,j}$ to denote the $j$th coordinate. 
$\mathbb F_k$ denotes a field of characteristic $k$. 
We call an element $e$ of a set family $\mathcal{F}$ as \emph{isolated} if there is no set of cardinality at least two containing $e$.
%\begin{definition}
%    For a given set system $\mathcal F\subseteq 2^U$ we say an element $e\in U$ is isolated if $\set{e}\subsetneq S$ implies $S\not \in \mathcal F$.
%\end{definition}

%\medskip \noindent \textbf{Packing $d$-dimensional rectangles. }
\paragraph{Packing $d$-dimensional rectangles}
As mentioned earlier, in this work, we only consider the setting where we do not allow rotations of the rectangles.
Thus, a set of $d$-dimensional rectangles\footnote{For the ease of notation rectangles without a fixed position are specified as vectors throughout the paper.} $\mathbf{v}_1, \mathbf{v}_2, \ldots, \mathbf{v}_k \in (0,1]^d$ where $\mathbf{v}_i = \{ v_{i,1}, v_{i,2}, \ldots, v_{i,d}\}, i \in [k]$ fits in the $d$-dimensional unit cube $\mathbf{1}^d$ if and only if there exists a positioning of these rectangles such that they all fit in $\mathbf{1}^d$, and they do not intersect with each other, i.e., there exist $\mathbf{p}_1, \mathbf{p}_2, \ldots, \mathbf{p}_k \in [0,1]^d$ where $\mathbf{p}_i = \{ p_{i,1}, p_{i,2}, \ldots, p_{i,d}\}, i \in [k]$ such that the following two conditions hold: 
\begin{enumerate}
	\item First, the rectangles fit inside the unit cube, i.e., for every $i \in [k], l \in [d]$, $ p_{i,l} + v_{i,l} \leq 1$. 
	\item The rectangles do not intersect with each other, i.e., the $k$ subsets of $[0,1]^d$ for $i \in [k]$ 
	\[
	[\mathbf{p}_i, \mathbf{p}_i + \mathbf{v}_i) := [p_{i,1}, p_{i,1}+v_{i,1}) \times 	[p_{i,2}, p_{i,2}+v_{i,2}) \times \dots \times 	[p_{i,d}, p_{i,d}+v_{i,d})
	\]
	are mutually disjoint. 
\end{enumerate} 
\section{Reduction from Graph Coloring}
\label{sec:proof}

In this section, we prove the inapproximability result for Geometric Bin Packing.
Our reduction is reminiscent of the reduction from graph coloring to vector bin packing by Chekuri and Khanna \cite{CK04} which showed there is no $d^{1/2-\epsilon}$ approximation for Vector Bin Packing unless $\NP=\ZPP$.

%\begin{figure}
%    \centering
%    %\includegraphics[width=30em]{clique1.jpeg}
%    \input{images/combined}
%    \caption{Embedding a graph according to \Cref{lem:coloringtogdp} and packing the cliques $\set{1,2},\set{3}$.}
%    \label{fig:embedding}
%\end{figure}

\medskip \noindent \textbf{Reduction.}
Our reduction outputs a $d$-dimensional Geometric Bin Packing instance from a given Graph Coloring instance $G=([n],E)$ wherein each $d$-dimensional rectangle corresponds to a vertex. 
The key idea is that the $d$-dimensional rectangles corresponding to a subset of vertices fit in the unit $d$-dimensional cube if and only if the subset of vertices is a clique in the graph. 
Fix a constant $\alpha \in (0,0.1)$.
We have $d=n$, and the $d$-dimensional rectangles $\mathbf{v}_1, \mathbf{v}_2, \ldots, \mathbf{v}_n$ are defined as follows. 
\[
v_{i,j} := \begin{cases}
\alpha,\text{ if }i=j \\
0.5+\alpha,\text{ if }(i,j) \in E\\
1,\text{ if }i \neq j, (i,j) \notin E \end{cases}
\]

Before analyzing the reduction, we need the following lemma regarding packing two $d$-dimensional rectangles in the unit $d$-dimensional cube $\mathbf{1}^d$. 
\begin{lemma} 
\label{lem:gpd-pair}
The $d$-dimensional rectangles $\mathbf u, \mathbf v \in [0,1]^d$ fit in the unit cube $\mathbf{1}^d$ if and only if there exists $j \in [d]$ such that $u_j + v_j \leq 1$. 
\end{lemma}
\begin{proof}
First, suppose that there exists $j \in [d]$ such that $u_j + v_j \leq 1$. We consider the following positions of the rectangles: $\mathbf{p}, \mathbf{q} \in [0,1]^d$: $p_l = 0 $ for all $l \in [d]$, and 
\[
q_l = \begin{cases}
0, \text{ if }l \neq j.\\
u_j, \text { if }l=j.
\end{cases}
\]
Note that $[\mathbf{p}, \mathbf{p}+\mathbf{u}) \cap [\mathbf{q}, \mathbf{q}+\mathbf{v}) =\emptyset$. 
%\newpage %hack to get footnote 7 on the same page.
Conversely, suppose that the two $d$-dimensional rectangles fit in a $d$-dimensional cube, i.e., there exist $\mathbf{p}, \mathbf{q}$ such that $[\mathbf{p}, \mathbf{p}+\mathbf{u}) \cap [\mathbf{q}, \mathbf{q}+\mathbf{v}) =\emptyset$. Note that $u_l + v_l > 1$ implies that $[p_l, p_l + u_l) \cap [q_l, q_l + v_l) \neq \emptyset$. Thus, if $u_l + v_l > 1$ for every $l \in [d]$, we get that $[\mathbf{p}, \mathbf{p}+\mathbf{u}) \cap [\mathbf{q}, \mathbf{q}+\mathbf{v}) \neq \emptyset$, a contradiction. Hence, there exists $l \in [d]$ such that $u_l + v_l \leq 1$. 
\end{proof}

\begin{figure}
    \centering
    \begin{subfigure}[b]{0.3\textwidth}
        \centering
        \includegraphics{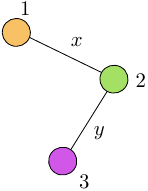}
        \caption{Input graph}
    \end{subfigure}
    \begin{subfigure}[b]{0.3\textwidth}
        \centering
        \includegraphics{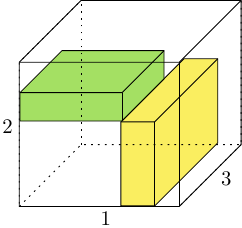}
        \caption{The clique $\set{1,2}$}
    \end{subfigure}
    \begin{subfigure}[b]{0.3\textwidth}
        \centering
        \includegraphics{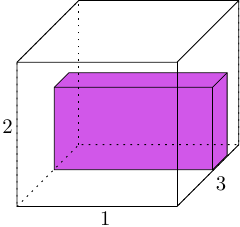}
        \caption{The clique $\set{3}$}
    \end{subfigure}
    
    \caption{Embedding a graph according to \autoref{lem:coloringtogdp} and packing the cliques $\set{1,2},\set{3}$.}
    \label{fig:embedding}
\end{figure}

We are now ready to analyze the reduction. 
\iffalse 
Each item in this instance corresponds to a vertex and each component is either 
The completeness is extremely simple.
For the soundness we use the observation\footnote{Notation: for $\mathbf a,\mathbf b\in \R^d$, $(\mathbf a,\mathbf b)=(a_1,b_1) \times (a_2,b_2) \dots \times (a_d,b_d)$.} that if $(\mathbf a,\mathbf b)\cap (\mathbf c,\mathbf d)=\emptyset$ then there exists $i$ such that $(a_i,b_i)\cap (c_i,d_i)=\emptyset$.
\fi

\begin{lemma}
\label{lem:coloringtogdp}
Given a graph $G=([n],E)$, $\alpha\in (0, 0.1)$, and the $d$-dimensional rectangles $\mathbf v_1, \mathbf v_2, \ldots, \mathbf v_n \in [0,1]^d$ defined as above with $d=n$, for every subset of vertices $S \subseteq [n]$, $S$  is a clique in $G$ if and only if $\{ \mathbf v_i: i \in S \}$ fit in $\mathbf{1}^d$. 
\end{lemma}

\begin{proof}
First, suppose that indices $i,j \in [n]$ are such that $(i,j)$ are not adjacent in $G$. By the definition of the vectors, for every $l \in [d]$, we have $v_{i,l}+v_{j,l}>1$. Thus, by~\autoref{lem:gpd-pair},  $\mathbf v_i, \mathbf v_j$ do not fit in $\mathbf{1}^d$. Thus, if a subset of vertices $S$ is not a clique, then the corresponding $d$-dimensional rectangles do not fit in the $d$-dimensional unit cube. 

Now, we define the positions $\mathbf p_1, \mathbf p_2, \ldots, \mathbf p_n \in [0,1]^d$ as follows: 
\[
p_{i,j} = \begin{cases}
0, \text { if } i \neq j \\
0.6,  \text { if } i = j.
\end{cases}
\]
Suppose that $(i,j) \in E$. Then, we have $[p_{i,i}, p_{i,i}+v_{i,i}) \cap [p_{j,i}, p_{j,i}+v_{j,i}) = \emptyset$. Thus, $[\mathbf p_i, \mathbf p_i + \mathbf v_i) \cap [\mathbf p_j, \mathbf p_j + \mathbf v_j) = \emptyset$. Hence, 
if a subset $S = \{ i_1, i_2, \ldots, i_s\}$ is a clique in $G$, then the cuboids $[\mathbf p_{i_1}, \mathbf p_{i_1}+\mathbf v_{i_1}), \ldots, [\mathbf p_{i_s}, \mathbf p_{i_s}+\mathbf v_{i_s})$ are all mutually disjoint. In other words, the set of rectangles $\mathbf v_{i_1}, \mathbf v_{i_2}, \ldots, \mathbf v_{i_s}$ fit in the $d$-dimensional unit cube. 
\end{proof}

Thus, the minimum number of cubes needed to cover all the rectangles is equal to the chromatic number of $\overline{G}$, i.e., the complement of $G$.
Note that here we are using the fact that the clique cover number of $G$ is equal to the chromatic number of $\overline{G}$.
Now, using the $n^{1-\epsilon}$ hardness for graph coloring by Zuckerman~\cite{Zuc07} and Feige and Kilian~\cite{FK98} we have the following result.
\main*
%Finally, as a consequence of the reduction in \autoref{lem:coloringtogdp} we can also conclude there is no $d^\delta$-approximation, for some $\delta>0$, for GBP under the weaker assumption of $\NP \ne \P$.
%This is because Lund and Yannakakis~\cite{LY94} showed there is no polynomial-time algorithm with an approximation ratio better than $n^\delta$, for some $\delta>0$, for graph coloring, unless $\P = \NP$. 
\section{Geometric Packing Dimension}
\label{sec:gpd}

We first formally define the geometric packing dimension $\textsf{gpd}(\mathcal{F})$ of a set family $\mathcal{F}$. 

\begin{definition}[Geometric Packing Dimension]
	For a set family $\mathcal{F}$ on a finite universe $U$, the geometric packing dimension $\gpd(\F)$ is the smallest integer $d$ such that there is an embedding $f: U \rightarrow [0,1]^d$ from $U$ to $d$-dimensional axis parallel rectangles such that for every subset $S \subseteq U, S = \{s_1, s_2, \ldots, s_t\}$, $S \in \F$ if and only if the set of $d$-dimensional rectangles $\{ f(s_1), f(s_2), \ldots, f(s_t)\}$ fit in $\mathbf{1}^d$. If no such $d$ exists, we say that $\gpd(\F)$ is infinite. 
\end{definition}

\subsection{GPD of downward closed set families}

Before proving~\autoref{thm:gpd-infinite}, we prove a couple of lemmas. 
First, we give a sufficient condition for three $d$-dimensional rectangles to fit inside the $d$-dimensional unit cube.

\begin{figure}
    \centering
    
    \begin{subfigure}[b]{0.45\textwidth}
        \centering
        \includegraphics{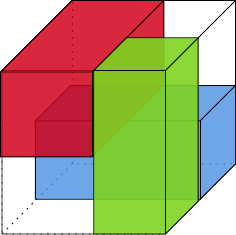}
        \caption{$|\set{j_{12},j_{23},j_{31}}|=3$}
    \end{subfigure}
    \begin{subfigure}[b]{0.45\textwidth}
        \centering
        \includegraphics{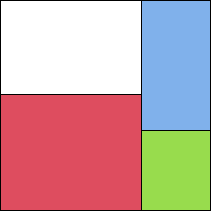}
        \caption{$|\set{j_{12},j_{23},j_{31}}|=2$}
    \end{subfigure}
    
    \caption{Packing of $\mathbf{v_1,v_2,v_3}$ from \autoref{lem:gpd-triple} projected on $j_{12},j_{23},j_{31}$.}
    \label{fig:threeitems}
\end{figure}

\begin{lemma}
\label{lem:gpd-triple}
Consider three $d$-dimensional rectangles $\mathbf{v}_1, \mathbf{v}_2, \mathbf{v}_3 \in [0,1]^d$, and suppose there exists $j_{12} \in [d]$ such that $v_{1,j_{12}} + v_{2,j_{12}} \leq 1$.
Also, suppose there exists $j_{23}, j_{31}  \in [d]$ such that $v_{1,j_{31}} + v_{3,j_{31}} \leq 1$, and $v_{2,j_{23}} + v_{3,j_{23}} \leq 1$.
If $\left|\{ j_{12}, j_{23}, j_{31} \} \right| \geq 2 $, then the three rectangles fit inside the unit $d$-dimensional cube. 
\end{lemma}
\begin{proof}
First, we consider the case when $\left|\{ j_{12}, j_{23}, j_{31} \} \right| = 3$, i.e., the three indices are all distinct. Then, we give the following positions $\mathbf{p}_1, \mathbf{p}_2, \mathbf{p}_3$: we set $p_{i,l}=0$ for each $i \in [3]$ and $l \notin \{ j_{12}, j_{23}, j_{31}\}$, and the rest of the values are set as follows. 

\begin{equation*}
	\begin{aligned}[c]
		p_{1,j_{12}} &= 0, \\ 
		p_{1,j_{23}} &= 0, \\ 
		p_{1,j_{31}} &= v_{3,j_{31}},  		
	\end{aligned}
	\qquad \qquad
	\begin{aligned}[c]
		p_{2,j_{12}} &= v_{1,j_{12}}, \\ 
		p_{2,j_{23}} &= 0, \\ 
		p_{2,j_{31}} &=   0,		
	\end{aligned}
	\qquad \qquad
\begin{aligned}[c]
	p_{3,j_{12}} &=  0,\\ 
	p_{3,j_{23}} &= v_{2,j_{23}}, \\ 
	p_{3,j_{31}} &=   0.		
\end{aligned}
\end{equation*}
With these parameters, we can observe that the subsets $[\mathbf{p}_i, \mathbf{p}_i +\mathbf{v}_i), i \in [3]$ are mutually disjoint. 

Next, we consider the case when $\left|\{ j_{12}, j_{23}, j_{31} \} \right| = 2$ (see \autoref{fig:threeitems} for an illustration). Without loss of generality, suppose that $j_{12}=j_{31}$. We give the following positions $\mathbf{p}_1, \mathbf{p}_2, \mathbf{p}_3$: we set $p_{i,l}=0$ for each $i \in [3]$ and $l \notin \{ j_{12}, j_{23}\}$, and the rest of the values are set as follows. 
\begin{equation*}
	\begin{aligned}[c]
		p_{1,j_{12}} &= 0 \\ 
		p_{1,j_{23}} &= 0 
	\end{aligned}
	\qquad \qquad
	\begin{aligned}[c]
		p_{2,j_{12}} &= v_{1,j_{12}} \\ 
		p_{2,j_{23}} &= 0 
	\end{aligned}
	\qquad \qquad
	\begin{aligned}[c]
		p_{3,j_{12}} &=   v_{1,j_{12}}\\ 
		p_{3,j_{23}} &= v_{2,j_{23}}  
	\end{aligned}
\end{equation*}
Similar to the above case, we have that the three subsets $[\mathbf{p}_i, \mathbf{p}_i +\mathbf{v}_i), i \in [3]$ are mutually disjoint. 
\end{proof}

Finally, we need~\autoref{lem:gpd-1} stated below. 
However, as the statement of this lemma is quite dense, it may be worthwhile to try to go through it at a high level.
Loosely speaking, it essentially says if, for some set of vectors (rectangles with unspecified positions), all coordinates except for one ($j$ in the lemma statement) are large, then only the small coordinate ($j$ in this case) decides if all of them fit in a bin or not. 
\begin{lemma}
\label{lem:gpd-1}
Suppose $\mathbf{v}_1, \mathbf{v}_2, \dots, \mathbf{v}_k \in [0,1]^d$ be a set of $d$-dimensional rectangles such that there exists $j \in [d]$ such that $v_{i_1,j} + v_{i_2,j} \leq 1$ for every $i_1, i_2 \in [k], i_1 \neq i_2$, and $v_{i_1,l} + v_{i_2,l} > 1$ for every $i_1, i_2 \in [k], i_1 \neq i_2, l \in [d], l \neq j$. Then, $\sum_{l\in [k]}v_{l,j} \leq 1$ if and only if the set of rectangles fit inside the $d$-dimensional unit cube. 
\end{lemma}

\begin{proof}
	Suppose that $\sum_{l\in [k]}v_{l,j} \leq 1$.
    Then, we give the following positions $\mathbf{p}_1, \mathbf{p}_2, \dots, \mathbf{p}_k$:
    we set $p_{i,l}=0$ for each $i \in [k]$ and $l \neq j$,
    set $p_{1,j} = 0$, and for $l\ne 1$ set
	\begin{align*}
		p_{l,j}=\sum_{i=1}^{l-1}v_{i,j}. 
	\end{align*}
Then, we have that the subsets $[\mathbf{p}_i, \mathbf{p}_i +\mathbf{v}_i), i \in [k]$ are mutually disjoint. 

Now, suppose that there exist positions $\mathbf{p}_1, \mathbf{p}_2, \dots \mathbf{p}_k$ such that the three subsets $[\mathbf{p}_i, \mathbf{p}_i +\mathbf{v}_i), i \in [k]$ are mutually disjoint.
Consider $i_1, i_2 \in [k], i_1 \neq i_2$. 
As $v_{i_1,l} + v_{i_2,l} > 1$ for every $ l \in [d], l \neq j$, for $[\mathbf{p}_{i_1}, \mathbf{p}_{i_1} +\mathbf{v}_{i_1})$ to be disjoint from $[\mathbf{p}_{i_2}, \mathbf{p}_{i_2} +\mathbf{v}_{i_2})$, we need that $[p_{i_1, j}, p_{i_1,j}+v_{i_1,j}) \cap [p_{i_2, j}, p_{i_2,j}+v_{i_2,j})  = \emptyset$. As this is true for every distinct pair of indices $i_1, i_2 \in [k]$, we obtain that $[p_{i,j}, p_{i,j}+v_{i,j}), i \in [k]$ are mutually disjoint, which proves that $\sum_{i=1}^k v_{i,j} \leq 1$. 
\end{proof}

We are now ready to prove that there are downward closed set systems without isolated elements that have infinite geometric packing dimension.
\gpdinfinite*

\begin{proof}
Fix an integer $n \geq 2$. Our counterexample is the following lines set system $\mathcal{F}$: 
\begin{enumerate}
\item The elements of the family are $U = \mathbb{F}_3^n$. 
\item  The set family consists of all the subsets of $U$ of size at most $2$, and sets of size $3$ of the form $\mathbf{u}, \mathbf{u}+\mathbf{v}, \mathbf{u}+2\mathbf{v}$, where $\mathbf{v} \neq \mathbf 0$ and $\mathbf{u}$ are elements of $U$, and the addition of vectors is the usual coordinatewise modulo $3$ addition of $\mathbb{F}_3$.   
\end{enumerate}
We claim that $\gpd(\F)$ is infinite. 

Suppose for contradiction that there is a mapping $f: U \rightarrow [0,1]^d$ from $U$ to $d$-dimensional rectangles that is a valid geometric packing dimension embedding. Consider an arbitrary element $a \in U$, and arbitrary elements $b_1, b_2 \in U \setminus \{a\}$. Let $i_1 \in [d] $ be such that $f(a)_{i_1} + f(b_1)_{i_1} \leq 1 $, and similarly define $i_2$. Note that such $i_1, i_2$ exist since $\{a,b_1\}, \{a,b_2\} \in \F$, and by~\autoref{lem:gpd-pair}.

First, we consider the case when $i_1 \neq i_2$. 
As there is a coordinate $j \in [d]$ where $f(b_1)_j + f(b_2)_j \leq 1$, we can infer that $\{a, b_1, b_2\} \in \mathcal{F}$ using~\autoref{lem:gpd-triple}. Now, consider an arbitrary element $b_3 \in U \setminus \{a, b_1, b_2 \}$. Note that there exists $i_3 \in [d]$ such that $f(a)_{i_3} + f(b_3)_{i_3} \leq 1$. However, there exists $\ell \in \{1,2\}$ such that $i_\ell \neq i_3$, implying that $a, b_3, b_\ell$ are a line, contradicting the fact that $a, b_1, b_2$ form a line in $\mathbb{F}_3^n$. 

Now, suppose that for every choice of $a, b_1, b_2 \in U$, no such distinct $i_1, i_2$ exist. Recall that for every pair of elements $a, b \in U$, there exists $l \in [d]$ such that  $f(a)_l + f(b)_l \leq 1$. 
The absence of such distinct $i_1, i_2$ implies that there is a single coordinate $j \in [d]$ such that for every pair of elements $a, b \in U$,  $f(a)_j + f(b)_j \leq 1$, and $f(a)_l + f(b)_l > 1$ for every $l \neq j$.
Now, using~\autoref{lem:gpd-1}, this implies that $f' : U \rightarrow [0,1]$ defined as $f'(u) = f(u)_j$ is a valid $1$-dimensional geometric embedding of $\mathcal{F}$. 

Finally, we prove that $\gpd (\F)  \neq 1$, finishing the proof. 
Pick 2 disjoint sets $S_1,S_2\in \mathcal F$ of size 3. Note that such sets are guaranteed to exist when $n \geq 2$. 
Observe that for any set $S\in \mathcal F$ of size 3 there exists $u\in S$ with $f'(u)\leq 1/3$ (by a simple averaging argument).
So, let $u_1\in S_1$ and $u_2\in S_2$ be such that $f'(u_1)\leq 1/3$ and $f'(u_2)\leq 1/3$.
Pick $d\in \mathbb F^n$ such that $d\not \in \set{0,u_2-u_1,u_1-u_2}$.
Now, consider the set $S_3=\set{u_1+d,u_2+d,2u_2-u_1+d}$.
Again, let $u_3\in S_3$ be such that $f'(u_3)\leq 1/3$.
Since $d\not \in \set{0,u_2-u_1,u_1-u_2}$, we have $u_3\not \in \set{u_1,u_2,2u_2-u_1}$ (observe that this is the only set in $\F$ containing both $u_1, u_2$).
Hence, $\set{u_1,u_2,u_3}\not\in \mathcal F$.
But $f'(u_1)+f'(u_2)+f'(u_3)\leq 1$, a contradiction.
\end{proof}

\subsection{GPD of bounded set systems}

In this subsection, we prove~\autoref{thm:gpd-simple}.
First, we define induced matching of a downward closed set system $\F$. 

\begin{definition}[Induced matching]
	Let $\mathcal{F} \subseteq 2^U$ be a downward closed set system. We say it has an induced matching of size $k$ if there exist $k$ mutually disjoint sets $U_1, U_2, \ldots, U_k \in \F$, each with cardinality at least two, such that for every non-empty set $S \in \F$, $S \cap U_i \neq \emptyset$ for at most one $i \in [k]$. %That is, $\F$ restricted to $\bigcup_{i \in [k]}U_i$ is a disjoint union of complete set systems, i.e., $\F \cap 2^{\bigcup_{i \in [k]}U_i} = \bigsqcup_{i \in [k]}2^{U_i}$. 
\end{definition}

We now show that the existence of a large induced matching implies that the geometric packing dimension of the set system is large as well. 

\begin{lemma}
\label{lem:matching}
	Suppose that $\F \subseteq 2^U$ has an induced matching of size $k$. Then, either $\gpd(\F)$ is infinite, or is at least $k$. 
\end{lemma}	

\begin{proof}
	As $\F$ has an induced matching of size $k$, there exist $k$ mutually disjoint sets $U_1, U_2, \ldots, U_k \in \F$ such that for every $S$ with $S \cap U_i \neq \emptyset, S \cap U_j \neq \emptyset$ for $i,j \in [k], i \neq j$, we have $S \notin \F$. Suppose for contradiction that $\gpd(\F) < k$, i.e., there exists a function $f : U \rightarrow [0,1]^d$ with $d < k$ that is a valid geometric packing. First, we consider $2k$ arbitrary distinct elements $a_1, a_2, \ldots, a_k , b_1, b_2, \ldots, b_k \in U$ such that $a_i, b_i \in U_i $ for every $i \in [k]$. As $\{ a_i, b_i \} \in \F$, using~\autoref{lem:gpd-pair}, we can conclude that there exists $s_i \in [d]$ such that $f(a_i)_{s_i}+f(b_i)_{s_i} \leq 1$. 
	
	We claim that $s_i \neq s_j$ for every $i,j \in [k], i \neq j$. Suppose for contradiction that there exists $i, j \in [k], i \neq j$ such that $s_i = s_j$. Note that we have $f(a_i)_{s_i}+f(b_i)_{s_i} \leq 1$, and $f(a_j)_{s_i}+f(b_j)_{s_i} \leq 1$.
    Therefore, at least one of $f(a_i)_{s_i}$, $f(b_i)_{s_i}$ is less than $1/2$ and at least one of $f(a_j)_{s_i}$, $f(b_j)_{s_i}$ is less than $1/2$.
    Without loss of generality, assume $f(a_i)_{s_i}\leq 1/2$ and $f(a_j)_{s_i}\leq 1/2$ (the other cases use the same arguments).
    So, we have $f(a_i)_{s_i} + f(a_j)_{s_i}\leq 1$ which along with \autoref{lem:gpd-pair} implies that $\{ a_i, a_j\} \in \F$.
    However, by the induced matching property, we must have that $\{ a_i, a_j\} \notin \F$, a contradiction.
    %Now, using~\Cref{lem:gpd-pair}, we get that $f(a_i)_{s_i}+f(a_j)_{s_i} >1 $. Similarly, we obtain that $f(a_i)_{s_i}+f(b_j)_{s_i} >1$, $f(b_i)_{s_i}+f(a_j)_{s_i} >1$ and $f(b_i)_{s_i}+f(b_j)_{s_i} >1 $.
\end{proof}

We are now ready to prove~\autoref{thm:gpd-simple}. 
\bounded*
\begin{proof}
Consider the graph $G=(V(G),E(G))$ defined as follows. 
\begin{enumerate}
    \item The vertex set $V(G)$ is the family of sets in $\mathcal F$ that have cardinality at least two.
    \item There is an edge between two (distinct) sets $S_1,S_2\in \mathcal F$ if there exists a set $S\in \mathcal F$ with $S \cap S_1 \neq \emptyset, S \cap S_2 \neq \emptyset$ ($S$ could be equal to either $S_1$ or $S_2$ as well).
\end{enumerate}
 As there are no isolated elements in $\mathcal{F}$ and there are at most $k$ elements in each set in $\mathcal F$, there are at least $\frac{|U|}{k}$ vertices in $G$.
Again, as each set in $\mathcal F$ has cardinality at most $k$ and each element appears in at most $B$ elements, the maximum degree of the graph $G$ is at most $(kB)^2$. 
Therefore, there must be an independent set $\mathcal M$ of size at least $\frac{| V(G)|}{(kB)^2+1}$ in $G$. By definition, the independent sets in $G$ are exactly the induced matchings in $\mathcal F$. Thus, there is an induced matching of size at least $\frac{|V(G)|}{(kB)^2+1}$ in $\mathcal F$.
Hence, by~\autoref{lem:matching} we get that $\gpd(\F)\geq \frac{|U|}{k(k^2B^2+1)}$.
\end{proof}

\section{Conclusion}
To summarize, in this paper, we could show a strong hardness of approximation result in the regime where $d$ is part of the input.
Then, in \autoref{thm:gpd-simple} and \autoref{thm:gpd-infinite} we showed that, in the fixed dimension regime, an extremely promising (and the only known) candidate reduction is unlikely to give a hardness of approximation result.

\section*{Acknowledgements}
We thank Arindam Khan and anonymous referees for their valuable comments.
We are especially grateful to the FSTTCS'23 referee, who found a bug in the proof of~\autoref{thm:gpd-infinite} presented in the previous versions, and to the IPL referees for pointing out the derandomization due to Zuckerman.

\bibliography{ref}

\end{document}